\documentclass[a4paper,twoside]{article}

\usepackage{epsfig}
\usepackage{subcaption}
\usepackage{calc}
\usepackage{amssymb}
\usepackage{amstext}
\usepackage{amsmath}
\usepackage{amsthm}
\usepackage{multicol}
\usepackage{pslatex}
\usepackage[bottom]{footmisc}
\usepackage[hidelinks]{hyperref}
\usepackage{natbib}
\usepackage{SCITEPRESS}     

\begin{document}

\title{SEQUENT: Towards Traceable Quantum Machine Learning using Sequential Quantum Enhanced Training \sup{*}}

\author{
    \authorname{
        Philipp Altmann\sup{1}, 
        Leo Sünkel\sup{1}, 
        Jonas Stein\sup{1}, 
        Tobias Müller\sup{2}, \\
        Christoph Roch\sup{1} and 
        Claudia Linnhoff-Popien\sup{1}
    }
    \affiliation{
        \sup{1}LMU Munich \\ \sup{2}SAP SE, Walldorf, Germany 
    }
    \email{philipp.altmann@ifi.lmu.de}
}



\abstract{ 
Applying new computing paradigms like quantum computing to the field of machine learning has recently gained attention. 
However, as high-dimensional real-world applications are not yet feasible to be solved using purely quantum hardware, hybrid methods using both classical and quantum machine learning paradigms have been proposed.
For instance, transfer learning methods have been shown to be successfully applicable to hybrid image classification tasks. 
Nevertheless, beneficial circuit architectures still need to be explored.
Therefore, tracing the impact of the chosen circuit architecture and parameterization is crucial for the development of beneficially applicable hybrid methods. 
However, current methods include processes where both parts are trained concurrently, therefore not allowing for a strict separability of classical and quantum impact. 
Thus, those architectures might produce models that yield a superior prediction accuracy whilst employing the least possible quantum impact.
To tackle this issue, we propose \textit{Sequential Quantum Enhanced Training} (SEQUENT) an improved architecture and training process for the traceable application of quantum computing methods to hybrid machine learning.
Furthermore, we provide formal evidence for the disadvantage of current methods and preliminary experimental results as a proof-of-concept for the applicability of SEQUENT. 
}

\onecolumn \maketitle \normalsize 
\renewcommand*{\thefootnote}{\fnsymbol{footnote}}
\footnotetext[1]{Citation: Altmann, P.; Sünkel, L.; Stein, J.; Müller, T.; Roch, C. and Linnhoff-Popien, C. (2023). SEQUENT: Towards Traceable Quantum Machine Learning Using Sequential Quantum Enhanced Training. In Proceedings of the 15th International Conference on Agents and Artificial Intelligence - Volume 3: ICAART, pages 744-751. DOI: \href{https://doi.org/10.5220/0011772400003393}{10.5220/0011772400003393}}
\setcounter{footnote}{0}
\renewcommand*{\thefootnote}{\arabic{footnote}}

\section{\uppercase{Introduction}}
\label{sec:introduction}
With classical computation evolving towards performance saturation, new computing paradigms like quantum computing arise, promising superior performance in complex problem domains. 
However, current architectures merely reach numbers of 100 quantum bits (qubits), prone to noise, and classical computers run out of resources simulating similar sized systems \citep{preskill2018quantum}. 
Thus, most real world applications are not yet feasible solely relying on quantum compute.
Especially in the field of machine learning, where parameter spaces sized upwards of 50 million are required for tasks like image classification, the resources of current quantum hardware or simulators is not yet sufficient for pure quantum approaches \citep{he2016deep}.
Therefore, hybrid approaches have been proposed, where the power of both classical and quantum computation are united for improved results \citep{bergholm2018pennylane}.
By this, it is possible to leverage the advantages of quantum computing for tasks with parameter spaces that cannot be computed solely by quantum computers due to hardware and simulation limitations.
Within those hybrid algorithms the quantum part is, analogue to the classical \textit{deep neural networks} (DNNs), represented by so called \textit{variational quantum circuits} (VQCs), which are parameterized and can be trained in a supervised manner using labeled data \citep{cerezo2021variational}. 
For hybrid machine learning, we will from hereon refer to VQCs as quantum parts and to DNNs as classical parts.

To solve large-scale real-world tasks, like image classification, the concept of \textit{transfer learning} has been applied for training such hybrid models \citep{girshick2014rich,pan2010}. 
Given a complex model, with high-dimensional input- and parameter spaces, the term transfer leaning classically refers to the two-step procedures of  first pre-training using a large but generic dataset and secondly fine-tuning using a smaller but more specific dataset \citep{torrey2010transfer}. 
Usually, a subset of the model's weights are frozen for the fine-tuning to compensate for insufficient amounts of fine-tuning data. 

Applied to hybrid \textit{quantum machine learning} (QML), the pre-trained model is used as a feature extractor and the dense classifier is replaced by a hybrid model referred to as \textit{dressed quantum circuit} (DQC) including classical pre- and post-processing layers, and the central VQC \citep{mari2020transfer}.  
This architecture results in concurrent updates to both classical and quantum weights. 
Even though, this produces updates towards overall optimal classification results, it does not allow for tracing the advantageousness of the quantum part of the architecture.
Thus, besides providing competitive classification results, such hybrid approaches do not allow for valid judgment whether the chosen quantum circuit benefits the classification.
The only arguable result is that it does not harm the overall performance or that the introduced inaccuracies may be compensated by the classical layers in the end.
However, as we currently are still only exploring VQCs, this verdict, i.e. traceability of the impact of both the quantum and the classical part, is crucial to infer the architecture quality from common metrics.
Overall, with current approaches we find a mismatch between the goal of exploring viable architectures and the process applied.

We therefore propose the application of \textit{Sequential Quantum Enhanced Training} (SEQUENT), an adapted architecture and training procedure for hybrid quantum transfer learning, where the effect of both classical and quantum parts are separably assessable.
We see our contributions as follows: 
\begin{itemize}
    \item We provide formal evidence that current quantum transfer learning architectures might result in an optimal network configuration (perfect classification / regression results) with the least-most quantum impact, i.e., a solution equivalent to a purely classical one.
    \item We propose SEQUENT, a two-step procedure of classical pre-training and quantum fine-tuning using an adapted architecture to reduce the number of features classically extracted to the number of features manageable by the VQC producing the final classification. 
    \item We show competitive results with a traceable impact of the chosen VQC on the overall performance using preliminary benchmark datasets.
\end{itemize}

\section{\uppercase{Background}}
To delimit SEQUENT, the following section provides a brief general introduction to the related fields of \hyperref[sec:quantum-computation]{quantum computation}, \hyperref[sec:qml]{quantum machine learning}, \hyperref[sec:deep-learning]{deep learning} and \hyperref[sec:transfer-learning]{transfer learning}. 

\subsection{Quantum Computing}\label{sec:quantum-computation}
\paragraph{Quantum Computation}
works fundamentally different than classical computation, since QC uses qubits instead of classical bits.
Where classical bit can be in the state 0 or 1, the corresponding state of a qubit is described in Dirac notation as $\mid0\rangle$ and $\mid1\rangle$. 
However, more importantly, qubits can be in a superposition, i.e., a linear combination of both:
\begin{equation}\label{eq:superposition}
\mid\psi\rangle = \alpha\mid0\rangle + \beta\mid1\rangle 
\end{equation}

To alter this state, a set of reversible unitary operations like rotations can be applied sequentially to individual \textit{target} qubits or in conjunction with a \textit{control} qubit.
Upon measurement, the superposition collapses and the qubit takes on either the state $\mid0\rangle$ or $\mid1\rangle$ according to a probability. 
Note that $\alpha$ and $\beta$ in  \eqref{eq:superposition} are complex numbers where $\mid\alpha\mid^2$ and $\mid\beta\mid^2$ give the probability of measuring the qubit in state $\mid0\rangle$ or $\mid1\rangle$ respectively. Note that $\mid\alpha\mid^2 + \mid\beta\mid^2 = 1$, i.e., the probabilities sum up to 1. \citep{nielsen2002quantum}

Quantum algorithms like Grover \citep{grover1996fast} or Shor \citep{shor1994algorithms} provide a theoretical speedup compared to classical algorithms. 
Moreover, in 2019 quantum supremacy was claimed \citep{arute2019quantum}, and the race to find more algorithms providing a quantum advantage is currently underway.
However, the current state of quantum computing is often referred to as the \textit{noisy-intermediate-scale quantum} (NISQ) era \citep{preskill2018quantum}, a period when relatively small and noisy quantum computers are available, however, still no error-correction to mitigate them, limiting the execution to small quantum circuits.
Furthermore, current quantum computers are not yet capable to execute algorithms that provide any quantum advantage in a practically useful setting. 
Thus, much research has recently been put into the investigation of hybrid-classical-quantum algorithms.
That is, algorithms that consist of quantum and classical parts, each responsible for a distinct task. 
In this regard, quantum machine learning has been gaining in popularity.

\paragraph{Quantum Machine Learning}\label{sec:qml}
algorithms have been proposed in several varieties over the last years \citep{farhi2014quantum,dong2008quantum,biamonte2017quantum}. 
Besides quantum kernel methods \citep{schuld2019quantum} \textit{variational quantum algorithms} (VQAs) seem to be the most relevant in the current NISQ-era for various reasons \citep{cerezo2021variational}. 

VQAs generally are comprised of multiple components, but the central part is the structure of the applied circuit or \textit{Ansatz}. 
Furthermore, a VQA Ansatz is intrinsically parameterized in order to use it as a predictive model by optimizing the parameterization towards a given objective, i.e. to minimize a given loss. 
Overall, given a set of data and targets, a parameterized circuit and an objective, an approximation of the generator underlying the data can be learned. 
Applying methods like gradient descent, this model can be trained to predict the label of unseen data \citep{cerezo2021variational,mitarai2018quantum}.
For the field of QML, various circuit architectures have been proposed \citep{biamonte2017quantum,khairy2020learning,schuld2020circuit}.

For the remainder of this paper, we consider the following \textit{simple} $\phi$-parameterized variational quantum circuit (VQC) for $\eta$ qubits: 
\begin{align}\label{eq:vqc}
\mathtt{VQC}_\phi(z) =  &\ \mathtt{meassure}_\sigma \circ \mathtt{entangle}_{\phi_\delta} \circ \dots \circ  \nonumber \\ 
                        &\ \circ \mathtt{entangle}_{\phi_1}\circ\mathtt{embed}_\eta(z)
\end{align}
with the depth $\delta$, and the output dimension $\sigma$ given the input $z = (z_1,\dots,z_\eta)$, where $\mathtt{embed}_\eta$ loads the data-points $z$ into $\eta$ balanced qubits in superposition via z-rotations, $\mathtt{entangle}_\phi$ applies controlled not gates to entangle neighboring qubits followed by $\phi$-parameterized z rotations, and $\mathtt{measure}_\sigma$ applies the Pauli-Z operator and measures the first $\sigma$ qubits \citep{schuld2019quantum,mitarai2018quantum}. 

This architecture has also been shown to be directly applicable to classification tasks, using the measurement expectation value as a one-hot encoded prediction of the target \citep{schuld2020circuit}.

Overall, VQAs have been shown to be applicable to a wide variety of classification tasks \citep{abohashima2020classification} and successfully utilized by \citet{mari2020transfer}, using the \textit{simple} architecture defined in \eqref{eq:vqc}.
Thus, to provide a proof-of-concept for SEQUENT, we will focus on said architecture for classification tasks and leave the optimization of embeddings \citep{larose2020robust} and architectures \citep{khairy2020learning} to future research. 

\subsection{Deep Learning}\label{sec:deep-learning}
\paragraph{Deep Neural Networks (DNNs)}
refer to parameterized networks consisting of a set of fully-connected layers.  
A layer comprises a set of distinct neurons, whereas each neuron takes a vector of inputs $x=(x_1,x_2,...x_n)$, which is multiplied with the corresponding weight vector $w_j=(w_j1,w_j2,...w_jn)$. A bias $b_j$ is added before being passed into an activation function $\varphi$. 
Therefore, the output of neuron $z$ at position $j$ takes the following form \citep{bishop2006pattern}: 
\begin{align}\label{eq:nn:layer}z_j = \varphi\left(\sum^n_{i=1}{w_{ji}x_{i} + b_j}\right)\end{align}

Given a target function $f(x): \mathbb{X} \mapsto y$, we can define the approximate
\begin{align}\label{eq:nn:f}
\hat{f}_\theta(x): \mathbb{X} \mapsto \hat{y} =L_{h_d\rightarrow o} \circ\dots\circ L_{n\rightarrow h_1}
\end{align}
as a composition of multiple layers $L$ with multiple neurons $z$ parameterized by $\theta$, $d-1$ $h$-dimensional hidden layers, and the respective input and target dimensions $n$ and $o$.
Using the prediction error $J = ( y - \hat{f}_\theta(x))^2$, $\hat{f}_\theta$ can be optimized by propagating the error backwards through the network using the gradient $\nabla_\theta J$ \citep{bishop2006pattern}.
Those feed forward models have been shown capable of approximating arbitrary functions, given a sufficient amount of data and either a sufficient depth (i.e. number of hidden layers) or width (i.e. size of hidden state) \citep{leshno1993multilayer}.
Deep neural networks for image classification tasks are comprised of two parts: A feature extractor containing a composite of convolutional layers to extract a $\upsilon$-sized vector of features $\mathtt{FE}: \mathbb{X} \mapsto \upsilon$, and a composite of fully connected layers to classify the extracted feature vector $\mathtt{FC}: \upsilon \mapsto\hat{y}$.
Thus, the overall model is defined as $ \hat{f}: \mathbb{X}\mapsto\hat{y} = \mathtt{FC}_\theta\circ\mathtt{FE}_\theta(x) $.
Those models have been successfully applied to a wide variety of real-world classification tasks \citep{he2016deep,imagenet2012}.
However, to find a parameterization that optimally separates the given dataset, a large amount of training data is required. 

\paragraph{Transfer Learning}\label{sec:transfer-learning}
aims to solve the problem of insufficient training data by transferring already learned knowledge (weights, biases) from a task $T_s$ of a source domain $D_s$ to a related target task $T_t$ of a target domain $D_t$. 
More specifically, a domain $D={\mathbb{X}, P(x)}$ comprises a feature space $\mathbb{X}$ and the probability distribution $P(x)$ where $x=(x_1,x_2,...,x_n) \in \mathbb{X}$. 
The corresponding task $T$ is given by $T = \{y,f(x)\}$ with label space $y$ and target function $f(x)$ \citep{zhuang2021comprehensive}.
A deep transfer learning task is defined by $\langle D_s,T_s,D_t,T_t,\hat{f}_t(\cdot) \rangle$, where $\hat{f}_t(\cdot)$ is defined according to \autoref{eq:nn:f} \citep{tan2018survey}.
Generally, transfer learning is a two-stage process. Initially, a source model is trained according to a specific task $T_s$ in the source domain $D_s$. Consequently, transfer learning aims to enhance the performance of the target predictive function $\hat{f}_t(\cdot)$ for the target learning task $T_t$ in target domain $D_t$ by transferring latent knowledge from $T_s$ in $D_s$, where $D_s \neq D_t$ and/or $T_s \neq T_t$. Usually, the size of $D_s >> D_t$ \citep{tan2018survey}. The knowledge transfer and learning step is commonly achieved via feature extraction and/or fine-tuning.

The \textbf{feature extraction} process freezes the source model and adds a new classifier to the output of the pre-trained model. Thereby, the feature maps learned from $T_s$ in $D_s$ can be repurposed and the newly-added classifier is trained according to the target task $T_t$ \citep{donahue2014decaf}.
The \textbf{fine-tuning} process additionally unfreezes top layers from the source model and jointly trains the unfreezed feature representations from the source model with the added classifier.
By this, the time and space complexity for the target task $T_t$ can be reduced by transferring and/or fine-tuning the already learned features of a pre-trained source model to a target model \citep{girshick2014rich}.

\section{\uppercase{Related Work}}
In the context of machine learning, VQAs are often applied to the problem of classification \citep{schuld2020circuit,mitarai2018quantum,havlivcek2019supervised,schuld2019quantum}, although other application areas exist. 
Different techniques, e.g. embedding \citep{lloyd2020quantum,larose2020robust}, or problems, e.g. barren plateaus \citep{mcclean2018barren}, have been widely discussed in the QML literature. However, we focus on hybrid quantum transfer learning \citep{mari2020transfer} in this paper.

Classical Transfer Learning is widely applied in present-day machine learning algorithms \citep{torrey2010transfer,pan2010,pratt1992} and can be extended with concepts of the emerging quantum computing technology \citep{Zen2020}. 
\citet{mari2020transfer} propose various hybrid transfer learning architectures ranging from classical to quantum (CQ), quantum to classical (QC) and quantum to quantum (QQ). 
The authors focus on the former CQ architecture, which which comprises the previously explained DQC.
In the current era of intermediate-scale quantum technology the DQC transfer learning approach is the most widely investigated and applied one, as it allows to some extend optimally pre-process high-dimensional data and afterwards load the most relevant features into a quantum computer. 
\citet{gokhale2020implementation} used this architecture to classify and detect image splicing forgeries, while \citet{acar2021covid} applied it to detect COVID-19 from CT images. 
Also, \citet{mari2020transfer} assess their approach exemplary on image classification tasks. 
Although the results are quite promising it is not clear from the evaluation, whether the dressed quantum circuit is advantageous over a fully classical approach.

\section{DQC QUANTUM IMPACT}\label{sec:dqc}

\begin{figure}[b]
\includegraphics[width=\linewidth]{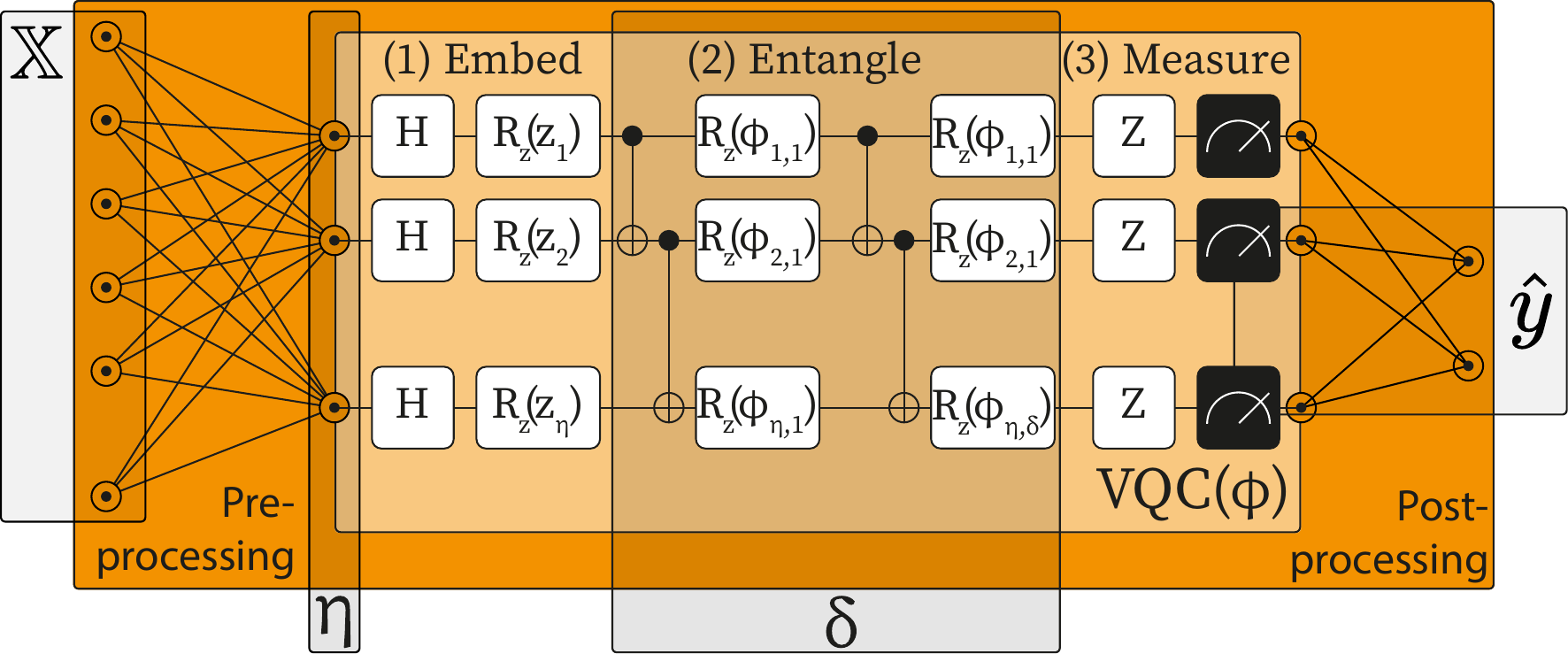}
  \caption{\textbf{Dressed Quantum Circuit Architecture}} 
  \label{fig:DQC}  
\end{figure}

We argue that within certain problem instance DQCs may yield accurate results while not making active use of any quantum effects in the VQC. 
This possibility exists especially for easy to solve problem instances, when all purely classical layers are sufficient to yield accurate results and the quantum layer represents the identity. 
This can be seen by realizing that the classical pre-processing layer acts as a hidden layer with a non-polynomial activation function, hence being capable of approximating arbitrary continuous functions depending on the number of hidden units by the universal approximation theorem \citep{leshno1993multilayer}.
Therefore, the overall DQC architecture is portrayed in \autoref{fig:DQC}.

The central VQC is defined according to \autoref{sec:qml} as introduced above.
Both pre- and post-processing layers are implemented by fully connected layers of neurons with a non-linear activation function according to \autoref{sec:deep-learning}.
Formally, the DQC for $\eta$ qubits can thus be depicted as:
\begin{align}\label{eq:DQC}
    \mathtt{DQC} = L_{\eta\rightarrow\sigma} \circ \mathtt{VQC}_\phi \circ L_{n\rightarrow\eta}
\end{align}
where $L_{n\rightarrow\eta}$ and $L_{\eta\rightarrow\sigma}$ are the fully connected classical \textit{dressing} layers according to \autoref{eq:nn:layer}, mapping from the input size $n$ to the number of qubits $\eta$ and from the number of qubits $\eta$ to the target size $\sigma$ respectively, and $ \mathtt{VQC}_\phi$ is the actual variational quantum circuit according to \autoref{eq:vqc} with $\eta$ qubits and $\sigma=\eta$ measured outputs.

Now let us consider a parameterization $\phi$, where $\mathtt{VQC}_\phi(z) = id(z) = z$ resembles the identity function. 
Consequently \eqref{eq:DQC} collapses into the following purely classical, 2-layer feed-forward network with the hidden dimension $\eta$:
\begin{align}\label{eq:universal}
    \mathtt{DQC} = L_{\eta\rightarrow\sigma} \circ id \circ L_{n\rightarrow\eta} =  L_{\eta\rightarrow\sigma} \circ L_{n\rightarrow\eta}
\end{align}

By the universal function approximation theorem, this allows $\mathtt{DQC}$ to approximate any polynomial function $f:\mathbb{R}^n\rightarrow{}\mathbb{R}^o$ of degree $1$ arbitrarily well, even if the VQC is not affecting the prediction at all. 

Consequently, one has to be careful in the selection of suitable problem instances, as they must not be too easy in order to ensure that the VQC is even needed to yield the desired results. This becomes especially difficult as current quantum hardware is quite limited, typically restricting the choice to fairly easy problem instances.
On top of this, no necessity to use a post-processing layer seems apparent, as it has been shown in various publications \citep{schuld2020circuit,schuld2019quantum} that variational quantum classifiers, i.e, VQCs can successfully complete classification tasks without any post-processing.
Overall, whilst conveying a proof-of-concept, that the combination of classical neural networks and variational quantum circuits in the dressed quantum circuit hybrid architecture is able to produce competitive results, this architecture is neither able to convey the advantageousness of the chosen quantum circuit nor exclude the possibility of the classical part just being able to compensate for quantum in-steadiness.

\section{SEQUENT}\label{sec:SEQUENT}
\begin{figure}[t!]
\includegraphics[width=\linewidth]{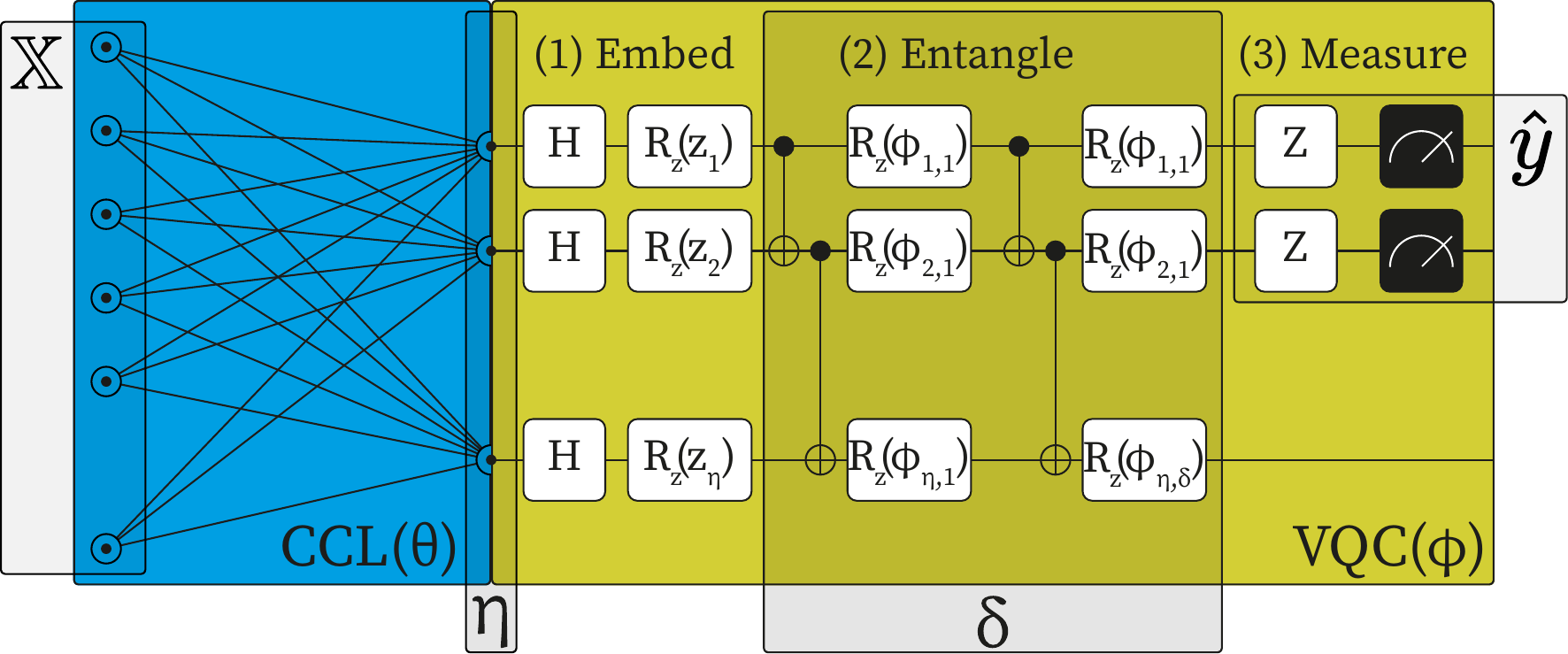}
  \caption{\textbf{SEQUENT Architecture}: Sequential Quantum Enhanced Training comprised of a classical compression layer (CCL) parameterized by $\theta$ and a variational quantum circuit (VQC) parameterized by $\phi$  with separate phases for classical (blue) and quantum (green) training for variable sets of input data  $\mathbb{X}$, prediction targets $\hat{y}$ and VQCs with $\eta$ qubits and $\delta$ entangling layers.} 
  \label{fig:SEQUENT}  
\end{figure}

To improve the traceability of quantum impact in hybrid architectures, we propose \textit{Sequential Quantum Enhanced Training}.
SEQUENT improves upon the dressed quantum circuit architecture by introducing two adaptations to it: 
\textbf{First}, we omit the classical post-processing layer and use the variational quantum circuit output directly as the classification result. 
Therefore we  reduce the measured outputs $\sigma$ from the number of qubits $\eta$ (cf. \autoref{fig:DQC}) to the dimension of the target $\hat{y}$ (cf. \autoref{fig:SEQUENT}).

The direct use of VQCs as a classifier has been frequently proposed and shown equally applicable as classical counterparts \citep{schuld2020circuit}.
By this, the overall quality of the chosen circuit and parameterization are directly assessable by the classification result, thus the final accuracy. 
Moreover, a parameter setting of universal approximation capabilities (cf. \autoref{eq:universal}) with the least (identitary) quantum contribution is mathematically precluded by the removal of the hidden state (compare \autoref{eq:DQC}).

Concurrently omitting the pre-processing or compression layer however would increase the number of at least required qubits to the number of output features of the problem domain, or, when applied to image classification, the chosen feature extractor (e.g. 512 for Resnet-18).
However, both current quantum hardware and simulators do not allow for arbitrate sized circuits, especially maxing out at around 100 qubits. 

\begin{figure}[b!]
\includegraphics[width=\linewidth]{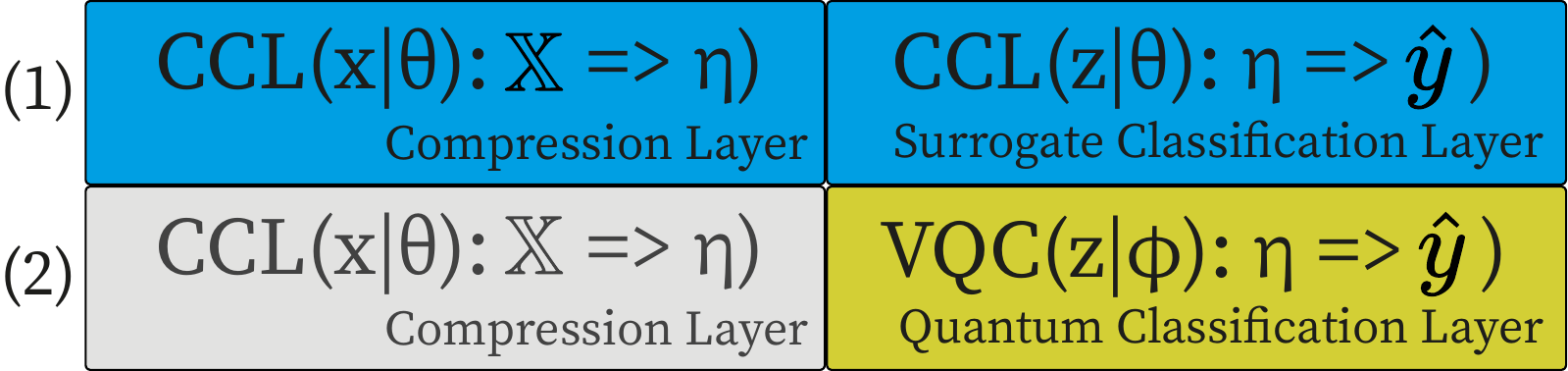}
  \caption{\textbf{SEQUENT Training Process} consisting of a classical (blue) pre-training phase (1) and a quantum (green) fine-tuning phase (2).}\label{fig:process}   
\end{figure}

\newpage
We therefore \textbf{secondly} propose to maintain the classical compression layer to provide a mapping/compression $\mathbb{X}\mapsto\eta$ and, in order to fully classically pre-train the compression layer, add a surrogate classical classification layer $\eta \mapsto \hat{y}$.
Replacing this surrogate classical classification layer with the chosen variational quantum circuit to be assessed and freezing the pre-trained weights of the classical compression layer then allows for a second, purely quantum training phase and yield the following sequential training procedure depicted in \autoref{fig:process}:  
\begin{enumerate}
\item Pre-train SEQUENT: $\hat{f}: \mathbb{X}\mapsto\eta\mapsto\hat{y} = \mathtt{CCL}_\theta(x) \circ  \mathtt{CCL}_\theta(z)$ containing a classical compression layer and a surrogate classification layer by optimizing the classical weights $\theta$
\item Freeze the classical weights $\theta$, replace the surrogate classical classification layer by the variational quantum classification circutit $\mathtt{VQC}_\phi $(cf. \autoref{eq:vqc}) and optimize the quantum weights $\phi$. 
\end{enumerate}

This two-step procedure can be seen as an application of transfer learning on its own, transferring from classical to quantum weights in a hybrid architecture. 

Overall, the SEQUENT architecture displayed in \autoref{fig:SEQUENT} can be formalized as: 
\begin{align}\label{eq:SEQUENT}
\mathtt{SEQUENT}_{\theta,\phi}: \mathbb{X}\mapsto\eta\mapsto\hat{y} &= \mathtt{VQC}_\phi(z) \circ  \mathtt{CCL}_\theta(x) \\
\mathtt{CCL}_\theta(x): \mathbb{X}\mapsto\eta &= L_{n\rightarrow\eta} \quad \mathrm{(cf. \ \autoref{eq:nn:layer})} \nonumber \\ 
\mathtt{VQC}_\phi(z): \eta\mapsto\hat{y} &\ \quad \quad \quad \quad \ \mathrm{(cf. \ \autoref{eq:vqc})} \nonumber
\end{align}


\newpage
To be used for the classification of high-dimensional data, like images, the input $x$ needs to be replaced by the intermediate output of an image recognition model $z$ (cf. \autoref{sec:deep-learning}). 
Combining both two-step transfer learning procedures, the following three-step procedure is yielded:
\begin{enumerate}
\item Classically pre-train a full classification model (e.g. Resnet~\citep{he2016deep}) $\hat{f}: \mathbb{X}\mapsto\upsilon\mapsto\hat{y} = \mathtt{FC}_\theta(z) \circ \mathtt{FE}_\theta(x)$ to a large generic dataset (compare \autoref{sec:deep-learning})
\item Freeze convolutional feature extraction layers $\mathtt{FE}$ and fine-tune fully-connected layers consisting of a compression layer and a surrogate classification layer $\mathtt{FE}: \upsilon\mapsto\eta\mapsto\hat{y} =  \mathtt{CCL}_\theta(z) \circ \mathtt{CCL}_\theta(x)$.
\item Freeze classical weights and replace surrogate classification layer with VQC to train the quantum weights $\phi$ of the hybrid model:\\ $\hat{f}_{\theta, \phi}: \mathbb{X}\mapsto\upsilon\mapsto\eta\mapsto\hat{y} = \mathtt{VQC}_\phi(z) \circ \mathtt{CCL}_\theta(x) \circ \mathtt{FE}$
\end{enumerate}
For a classification task with $n$ classes, at least $\eta \geq n$ qubits are required.
Whilst we use the \textit{simple} Ansatz introduced in \autoref{eq:vqc} with $\eta=6$ qubits and a circuit depth of $\delta=10$ to validate our approach in the following, any VQC architecture yielding a direct classification result would be conceivable.

\section{EVALUATION}

\begin{figure}[t]
\includegraphics[width=\linewidth]{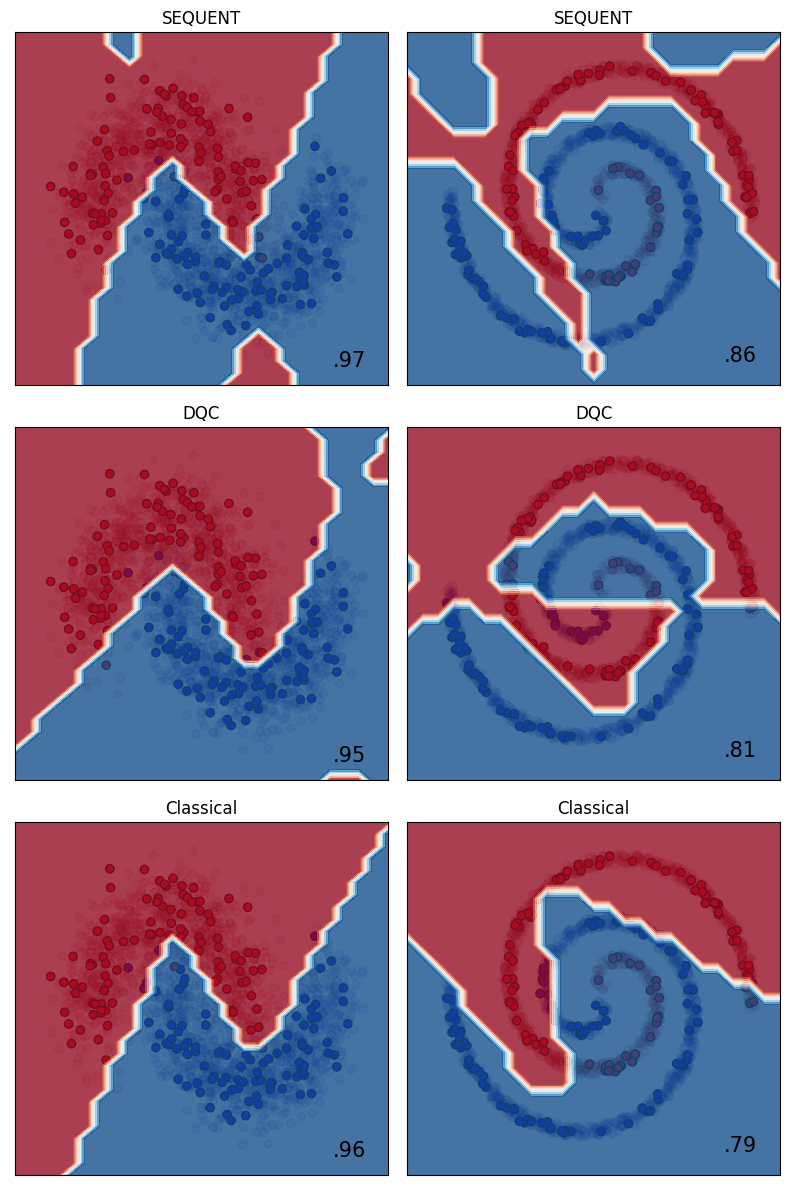}
  \caption{\textbf{Classification Results} of SEQUENT, DQC and Classical Feed Forward Neural Network for moons (left) and spirals (right) benchmark datasets}\label{fig:experiment}   
\end{figure}

We evaluate SEQUENT by comparing its performance to its predecessor, the DQC, and a purely classical feed forward neural network.
All models were trained on 2000 datapoints of the moons and spirals \citep{lang1988learning} benchmark dataset for two and four epochs of sequential, hybrid and classical training respectively. 
To guarantee comparability, we set the size of the hidden state of the classical model to $h = \eta = 6$.
The code for all experiments is available here\footnote{\url{https://github.com/philippaltmann/SEQUENT}}.
The classification results are visualized in \autoref{fig:experiment}.
Looking at the result for the moons dataset, all compared models are able to depict the shape underlying data. 
Note, that even the considerably simpler classical model is perfectly able to separate the given classes. 
Hence, these experimental results support the concerns about the impact of the VQC to the overall DQC's performance (cf. \autoref{sec:dqc}). 
With a final test accuracy of 95\%, the DQC performs even worse than the purely classical model reaching 96\%.
Looking at the SEQUENT results however, these concerns are eliminated, as the performance and final accuracy of 97\%, besides outperforming both compared models, can certainly be denoted to VQC, due to the applied training process and the used architecture. 
Similar results show for the second benchmark dataset of intertwined spirals on the right side of \autoref{fig:experiment}. 
The overall best accuracy of 86\% however suggests, that further adjustments to the VQC could be beneficial. 
This result also depicts the application of SEQUENT we imagine for benchmarking and optimizing VQC architectures.

\section{\uppercase{Conclusions}} \label{sec:conclusion}
We proposed \textit{Sequential Quantum Enhanced Training} (SEQUENT), a two-step transfer learning procedure applied to training hybrid QML algorithms combined with an adapted hybrid architecture to allow for tracing both the classical and quantum impact on the overall performance. 
Furthermore, we showed the need for said adaptions by formally pointing out weaknesses of the DQC, the current state-of-the-art approach to this regard. 
Finally, we showed that SEQUENT yields competitive results for two representative benchmark datasets compared to DQCs and classical neural networks.
Thus, we a provided proof-of-concept for both the proposed reduced architecture and the adapted transfer learning training procedure.  

However, whilst SEQUENT theoretically is applicable to any kind of VQC, we only considered the simple architecture with fixed angle embeddings and $\delta$ entangling layers as proposed by \citep{mari2020transfer}.
Furthermore, we only supplied preliminary experimental implications and did not yet test any high dimensional real-world applications.
Overall, we do not expect superior results that outperform state-of-the-art approaches in the first place, as viable circuit architectures for quantum machine learning are still an active and fast-moving field of research. 

Thus, both the real world applicability and the development of circuit architectures that indeed offer a benefit over classical ones should undergo further research attention. 
To empower real-world applications, the use of hybrid quantum methods should also be kept in mind when pre-training large classification models like Resnet. 
Also, applying more advanced techniques to train the pre-processing or compression layer to take full advantage of the chosen quantum circuit should be examined. 
Therefore, auto-encoder architectures might be applicable to train a more generalized mapping from the classical input-space to the quantum-space.
Overall, we belief, that applying the proposed concepts and building upon SEQUENT, both valuable hybrid applications and beneficial quantum circuit architectures can be found. 

\section*{\uppercase{Acknowledgements}}
This work is part of the Munich Quantum Valley, which is supported by the Bavarian state government with funds from the Hightech Agenda Bayern Plus and was partially funded by the German BMWK Project PlanQK (01MK20005I). 
\bibliographystyle{apalike}
{\small \bibliography{main.bib}}

\end{document}